\documentclass[twocolumn,showpacs,preprintnumbers]{revtex4}
\usepackage[dvips]{graphicx}

\begin{document}

\author{A.M. Bratkovsky$^{1}$ and A.P. Levanyuk$^{1,2}$}
\affiliation{$^{1}$ Hewlett-Packard Laboratories, 1501 Page Mill
Road, Palo
Alto, California 94304\\
$^{2}$Departamento de F\'{i}sica de la Materia Condensada, C-III,
Universidad Aut\'{o}noma de Madrid, 28049 Madrid, Spain }
\title{ Depolarizing field and ``real" hysteresis loops
in nanometer-scale ferroelectric films
}
\date{August 7, 2006}

\begin{abstract}
{ We give detailed analysis of the effect of depolarizing field in
nanometer-size ferroelectric capacitors studied by Kim {\it et al.}
[Phys. Rev. Lett. {\bf 95}, 237602 (2005)]. We calculate a
critical thickness of the homogeneous state and its stability with
respect to domain formation for strained thin films of BaTiO$_3$ on
SrRuO$_3$/SrTiO$_3$ substrate within the Landau theory. While the former (2.5nm)
is the same as given by ab-initio calculations, the actual critical
thickness is set by the domains at 1.6nm. There is a large Merz's
activation field for polarization relaxation. Remarkably, the
results show a {\em negative} slope of the ``actual" hysteresis
loops, a hallmark of the domain structures in ideal thin films with
imperfect screening.}
\end{abstract}
 \maketitle

Noh {\it et al}.\cite{KimL05,KimAPL05,nohNodl06} have recently published a
series of seminal experimental studies of very thin, down to $5$nm,
ferroelectric (FE) BaTiO$_{3}$ capacitors. They obtained hysteresis loops at
frequencies up to 100kHz and studied retention of a single domain (SD) state
in various external fields. They have applied high external field to obtain
polarization saturated state, perhaps a SD one, and observed its relaxation
when the external field $E_{0}$ were reduced below a certain value $%
E_{0}=E_{0r}$. Noh {\it et al}. have approximately identified this field as
a depolarizing field due to incomplete screening by electrodes and claimed
that it coincides with the one estimated from electrostatics. We show below
that such an interpretation does not apply and present a consistent
interpretation, which reveals important features of domain structure in
electroded thin FE films: (i)\ we find deviations from the Merz's empirical
formula\cite{Merz56} for thickness and exposition time dependence of the
activation field for domain wall movement and/or domain nucleation in very
thin films; (ii) replotting the hysteresis loops as function of a{\em \ }%
field in the ferroelectric reveals a {\em negative} susceptibility of
multidomain films governed by the electrostatics, predicted some time ago
\cite{BLPRB01}; (iii) finally, we argue that domain structure in the
thinnest ferroelectric films with electrodes may be different from that in
thick films.

In their study, Kim {\it et al}.\cite{KimL05} followed the idea by Mehta
{\em et al.}\cite{mehta73} that the incomplete screening of the
ferroelectric bound charge by electrodes leads to a depolarizing field
inside the ferroelectric that promotes domain nucleation and movement and
limits the polarization retention. Kim {\em et al.} speculated that the
depolarizing film can be identified with an external field $E_{0}=E_{0r}$
that stops the polarization relaxation. They claimed that $E_{0r}$ is very
close to the depolarizing field $E_{d}$. This is incorrect, however, since $%
E_{d}$ was estimated under {\em finite} external field $E_{0}$ from formula
by Mehta {\it et al}. who did only the case of short-circuited capacitor ($%
E_{0}=0$). If they were right, the Merz activation field for the domain
motion and/or nucleation in their ultrathin samples would have been
negligible, which is not the case (see below).

It is easy to find the external field $E_{0}=E_{0b}$ that {\em completely}
compensates the depolarization field (i.e. corresponds to zero field in FE).
This is the point where the field in FE changes sign from negative to
positive with regards to the polarization. The homogeneous\ field in the
monodomain ferroelectric is \cite{BLcm06}
\begin{eqnarray}
E_{f} &=&\left( E_{0}-2P\lambda /\epsilon _{0}\epsilon _{e}l\right) /\left(
1+2\lambda /\epsilon _{e}l\right) .  \label{eq:Efh} \\
&=&E_{0}+E_{d},  \nonumber
\end{eqnarray}
where $E_{0}=U/l$ is the external and
\begin{equation}
E_{d}=-PL_{0}/\epsilon _{0}l,\qquad L_{0}=2\lambda /\epsilon _{s}
\label{depfield}
\end{equation}
the depolarizing field, $L_{0}$ the characteristic length scale in
electrodes, and we have used the fact that in a metal the screening length
is small, $\lambda /\epsilon _{e}l\ll 1.$ The relation between the
polarization and the external field $E_{0}$ is then, using the equation of
state \cite{LiCross05} renormalized by strain \cite{pertsev98}:
\begin{equation}
A_{h}P+BP^{3}+CP^{5}+FP^{7}=\frac{E_{0}}{1+2\lambda /\epsilon _{e}l},
\label{eq:P}
\end{equation}
\begin{equation}
A_{h}=\hat{A}+\frac{2\lambda }{\epsilon _{0}(\epsilon _{e}l+2\lambda )}%
\simeq \hat{A}+\frac{2\lambda }{\epsilon _{0}\epsilon _{e}l}.  \label{eq:Ah}
\end{equation}
Here we have taken into account that the first coefficient may be
renormalized by the additional boundary conditions (ABC), $\hat{A}%
=A+(2\alpha +\beta )/l$ \cite{BL05}, but in the present case the ABC effect
is not important, since the variation of the spontaneous polarization on
thickness, where the $(2\alpha +\beta )/l$ term enters, is small, see
Fig.~1a (inset). The equation of state for spontaneous polarization $P_{s}$
takes the form
\begin{equation}
\hat{A}+BP_{s}^{2}+CP_{s}^{4}+FP_{s}^{6}=0,  \label{eq:PsABC}
\end{equation}
which is readily solved analytically. Since we know from the data the
polarization at zero external field, $P_{0}=P(E_{0}=0)$\cite{KimL05}, we can
indeed easily find $P_{s}$ from (\ref{eq:P}),(\ref{eq:PsABC}). We see from
Eq.~(\ref{eq:Efh})\ that $E_{f}=0$ when $E_{0b}=2\lambda P_{s}/\left(
\epsilon _{0}\epsilon _{e}l\right) $, the external field for
multidomain-single domain boundary.
\begin{figure}[tbp]
\includegraphics{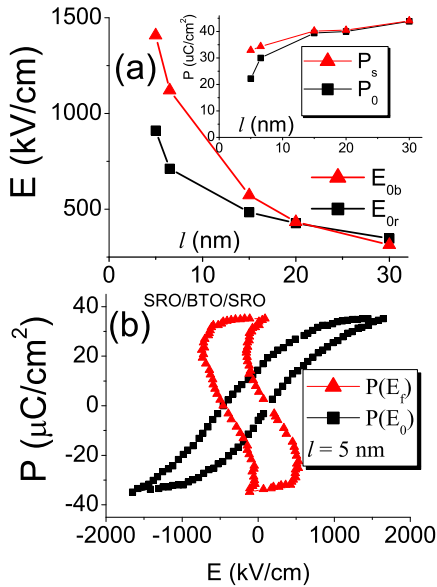}
\vskip -0.5mm \caption{ (a) The external field $E_{0b}$ (where
$E_f=0$) and $E_{0r}$ where the relaxation of polarization starts in
5~nm thick film\protect\cite{KimL05}. Inset: the spontaneous
polarization $P_s$ and the extrapolated $P_0$\protect\cite {KimL05}.
(b) The measured $P(E_0)$ and the ``actual" $P(E_f)$ hysteresis
loops. } \label{fig:fig1}
\end{figure}

Finite depolarizing field does not imply that the domains will be detected
after a certain observation time. Indeed, $E_{0b}=1450$kV/cm in $5$~nm
sample, which is about{\em \ }60\%{\em \ larger} than the relaxation field $%
E_{0r}$ identified by Kim {\it et al}. This means that during the
observation time $t_{relax}=10^{3}$s the domains begin to form only when
there is a field $E_{f}=-\left( 490\pm 70\right) ~$kV/cm opposite to the
polarization (at $E_{0}=E_{0r}=910$kV/cm for 5nm sample, raw data from Kim
{\it et al.}\cite{KimL05}). The activation field strongly depends on FE film
thickness for long application time ($\sim 10^{3}$s). Indeed, in the $l=30$
nm FE\ film the activation field is practically zero, Fig.1a. It is
interesting to see how the activation field depends on the application time.
To this end, we have replotted the hysteresis loops taken at 2~kHz as a
function of a field in the ferroelectric $E_{f}$, $P=P(E_{f}),$ Figs.1b,2
Note that we used Eq.~(\ref{depfield}) to calculate $E_{f},$ which applies
to homogeneous state. One, however, can apply it to an MD\ part of the loops
around $P\approx 0$ too, since the domains are narrow and the field in the
bulk is approximately homogeneous. Even in 5nm film the domain width $a=2.2$%
nm (see below), so this approximation should be semiquantitative. The use of
Eq.~(\ref{depfield}) is justified for finding the activation field for
domains (the ends of a horizontal parts of the loop). We see that the
activation field of similar magnitude\ is observed at all thicknesses. This
field is about the same for 5 nm sample as observed for much longer $10^{3}$%
s of application time.
\begin{figure}[tbp]
\includegraphics{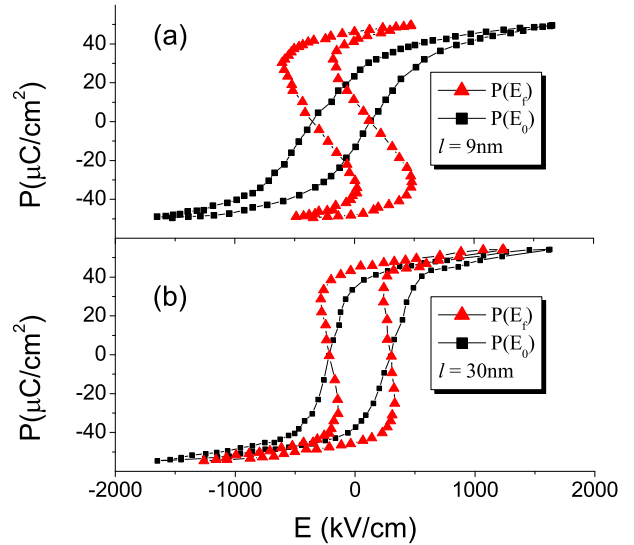}
\vskip -0.5mm \caption{ The measured $P(E_0)$ and the ``actual"
$P(E_f)$ hysteresis loops: (a) film thickness $l=9$~nm, (b)
$l=$30~nm.} \label{fig:fig2}
\end{figure}

Comparing the data by Kim {\it et al} with Merz \cite{Merz56}, one expects
that the switching time is close to the application time mentioned above. We
rewrite the Merz's empirical formula in the form
\begin{equation}
E_{a}=(\alpha \ln \tau )/l,  \label{eq:Emerz}
\end{equation}
where $\tau $ is the switching time and $\alpha $ the coefficient. It
describes a weak dependence of the activation field on the application time
for $l=5$nm as well as strong reduction of the field with a sample thickness
at a large application time rather well. However, weak dependence of $E_{a}$
on the film thickness in the case of small application times is not captured
by (\ref{eq:Emerz}).

We find using Eqs.~(\ref{eq:Efh})-(\ref{eq:P}) that the electric
field in short-circuited 5nm sample is $E_{f}(E_{0}=0)=-1200$~kV/cm,
exceeding the magnitude of the estimated activation field. This
means that in a short-circuited sample SD state relaxes quicker than
in $10^{3}$s. If the value of the activation field is defined by the
thickness only and not by properties of electrodes or an
electrode-film interface, one can speculate about the properties of
electrodes, which can facilitate a smaller field in\ a
short-circuited sample and a longer, at least $10^{3}$s, retention
of a SD state. We have found that for $d=5$nm such an electrode
should have $\lambda /\epsilon _{e}<0.033$\AA . Since in
\cite{KimL05} this value is about $0.1$\AA , it does not seem
impossible to find such an electrode. Alternatively, thinner films
may show longer retention, as Eq.~(\ref{eq:Emerz}) suggests.

An unusual specific feature of replotted loops is that they all have a {\em %
negative slope}, most pronounced at $l=5$nm. The exact value is affected by
error bars, but there is a strong indication that it is characteristic of
all samples. The negative slope has been predicted some time ago for an
ideal ferroelectric plate between perfect metallic electrodes with a voltage
drop across thin dead layers\cite{BLPRB01}: it is a hallmark of domain
structure governed mainly by electrostatics. There are no dead layers in the
present films \cite{nohNodl06}, but the voltage drop happens across a
screening layer with the thickness $\lambda $ in an electrode, with an
identical result and apparently beyond the precision of Ref.~\cite{nohNodl06}%
. Comparing theory with the data, the expression for the dielectric
constant, given by Eq.~(31) of Ref. \cite{BLPRB01}, can be simplified to
\begin{equation}
\epsilon _{f}\approx -\epsilon _{e}l/\lambda ,  \label{eq:negef}
\end{equation}
where $\epsilon _{f}=1+\epsilon _{0}^{-1}dP/dE_{f}|_{E_{f}=0}$ \
Substituting the numbers, we find the theoretical value $\epsilon _{f}=-525$
for equilibrium conditions while the experimental one found from Fig.1b (raw
data for $2$kHz, Ref.\cite{KimAPL05}) is $\epsilon _{f}=-680$, i.e they are
pretty close. According to Eq.~(\ref{eq:negef}), the absolute value of the
negative slope for $l=9$nm should be about two times larger than for $l=5$%
nm, and one sees from the Figs.1b,2 that this is indeed the case. If \ this
agreement is not accidental, it indicates that in the films studied by Kim
{\it et al}. the properties of the domain structure may be mainly defined by
the electrostatics and not by some specific structural features of the
samples, etc. The negative slope of the hysteresis loops is apparently
observed for the first time.

To reveal specific features of domain structure in thinnest films we take
into account that at small enough thickness the system should be in the
paraelectric phase (as shown explicitly below) and consider loss of
stability of this phase when the film thickness increases, i.e. a
paraelectric-FE phase transition{\em \ with thickness}. The loss of
stability is signalled by appearance of a non-trivial solution of the
equations of state which can be either homogeneous (SD) or inhomogeneous
(MD). A homogeneous solution of (\ref{eq:P}) is possible at $l>l_{h},$ where
$l_{h}=L_{0}/\epsilon _{0}|A,$ at room temperature $l_{h}^{\text{RT}}=3.5$%
nm. The domain instability means the appearance of a solution in form of a
``polarization wave'' $P=P_{k}(z)e^{ikx}$ \cite{Chensky82,BLinh} of the
(linearized) equation of state with the gradient term included:
\begin{equation}
AP-g\nabla _{\perp }^{2}P=E,  \label{eq:eqins}
\end{equation}
where $\nabla _{\perp }^{2}=\partial _{x}^{2}+\partial _{y}^{2}$ is
``in-plane'' Laplacian. In the case of metallic screening, this gives the
following condition\cite{BLcm06}:
\begin{equation}
\chi \tan \frac{1}{2}\chi kd=\epsilon _{\perp }k/\epsilon _{e}\sqrt{%
k^{2}+\lambda ^{-2}},  \label{eq:chi}
\end{equation}
where $\chi ^{2}=-\epsilon _{0}\epsilon _{\perp }(A+gk^{2})>0,$ $\epsilon
_{\perp }$ is the dielectric constant in the direction perpendicular to
ferroelectric axis in the plane of the film. The case of interest to us is $%
k\lambda \ll 1,$ easily met for metallic electrodes. We assume (and check
validity later)\ that $\epsilon _{\perp }\lambda k/\epsilon _{e}\chi \gtrsim
1.$ Then, the equation simplifies to $\chi kd=\pi ,$ the same as in FE\ film
without electrodes or with a dead layer. We then find the maximal value (the
highest temperature)\ of $A_{d}=-2gk_{c}^{2}$ at $k=k_{c}$ where this
equality is first met and domains begin to form:
\begin{equation}
-A_{d}=2gk_{c}^{2}=\xi /\epsilon _{0}l,\qquad k_{c}=\left( \pi ^{2}/\epsilon
_{\perp }\epsilon _{0}gl^{2}\right) ^{1/4},  \label{eq:Ad}
\end{equation}
where $\xi =2\pi \sqrt{g\epsilon _{0}/\epsilon _{\perp }}$ is the
characteristic length scale. Now, checking the assumption that we
used to solve the Eq.~(\ref{eq:chi}), we see that it boils down to
$\lambda \epsilon _{\perp }^{1/2}/\epsilon _{e}\left( \epsilon
_{0}g\right) ^{1/2}\gtrsim 1$. Using values of $\lambda ,\epsilon
_{e}$ from Ref. \cite{KimL05}, the value of $g$ from
Ref$.$\cite{shirane2}, and calculating $\epsilon _{\perp }$using the
coefficients of Ref. \cite{LiCross05}, we find that this condition
is satisfied. From Eq.~(\ref{eq:Ad}), we obtain the following
critical thickness for domains at room temperature:
\begin{equation}
l_{d}^{\text{RT}}=\xi /|A_{d}|\epsilon _{0}\simeq 3.1\text{ nm.}
\end{equation}
Since $l_{d}^{\text{RT}}<l_{h}^{\text{RT}}$, the phase transition is
into a MD state. The spatial distribution of a spontaneous
polarization is near sinusoidal at $l\gtrsim $ $l_{d}^{\text{RT}}$.
Higher harmonics develop with increasing thickness, and the
polarization distribution tends to a conventional structure with
narrow domain walls. But at small thicknesses it is nearly
sinusoidal, and one can expect weaker pinning compared to thicker
films. It is hardly surprising that the empirical Merz's formula
obtained for conventional domain structure does not apply to a
sinusoidal one. The
half-period of the sinusoidal domain structure can be estimated as $a_{c}^{%
\text{RT}}\approx \pi /k_{c}=1.7$ nm at the transition and as $a^{\text{RT}%
}=2.2$nm for $l=5$nm.

It is instructive to consider the phase transition with thickness at zero
Kelvin, where we get $\epsilon _{\perp }=408,$ $\xi =0.08$\AA , $l_{d}^{(0%
\text{K)}}=1.6$nm, $l_{h}^{(0\text{K)}}=2.5$nm. The last result (homogeneous
critical thickness of $2.5$nm) is remarkable, since it practically{\em \
coincides }with the ab-initio calculation for the critical thickness of $2.4$%
nm in Ref.\cite{ghosez03}. The ground state of the film is, however, not
homogeneous but multidomain, and the domain ferroelectricity appears in
films thicker than $l_{d}^{(0\text{K)}}=1.6$nm, which is the {\em true
``critical size''}\ for ferroelectricity in FE films in the present study.

We thank T.W. Noh and his group for kindly sharing their data and useful
discussions. APL is partially supported by MAT2003-02600 and
S-0505/MAT/000194.

\end{document}